\documentclass[3p]{elsarticle}

\usepackage{fancyhdr}
\usepackage{booktabs}
\usepackage{ctable}
\usepackage{amsmath}
\usepackage{mathtools}
\usepackage{commath}
\usepackage{graphicx}
\usepackage{nomencl}
\usepackage{commath}
\usepackage{natbib}
\usepackage{multirow}
\usepackage{longtable}
\usepackage{textcomp}
\usepackage{array}
\usepackage{graphics}
\usepackage{tabularx}
\usepackage{color}
\usepackage{geometry}
\usepackage{natbib}
\usepackage{mciteplus}
\usepackage{footnote}
\usepackage{amssymb}
\usepackage{pdflscape}
\usepackage{afterpage}
\usepackage{multirow}
\usepackage[title]{appendix}
\usepackage{subfigure}
\usepackage{caption}
\usepackage{amssymb}
\usepackage{natbib}
\usepackage{mleftright} 
\usepackage{latexsym}
\usepackage{color,soul}

\usepackage{etoolbox}
\renewcommand\nomgroup[1]{%
	\item[\bfseries
	\ifstrequal{#1}{A}{}{%
		\ifstrequal{#1}{G}{Greek Letters}{%
			\ifstrequal{#1}{S}{Subscripts}{}}}%
	]}

\usepackage[version=3]{mhchem} 

\journal{Energy \& Fuels
 }
\biboptions{numbers,sort&compress}

\begin{document}
\begin{frontmatter}

\title{Review of microscale dynamics of dilution\textendash induced asphaltene precipitation under controlled mixing conditions}

\author{Jia Meng\textsuperscript}
\author{Somasekhara Goud Sontti\textsuperscript}
\author{Xuehua Zhang\textsuperscript{*}\corref{cor1}}
\ead{xuehua.zhang@ualberta.ca}
\address{Department of Chemical and Materials Engineering, University of Alberta, Alberta T6G 1H9, Canada}

\newpage
\begin{abstract}	

	 As the most complex and heaviest component in bitumen, asphaltene precipitation induced by solvent dilution is important in oil sands extraction to remove solids and water from bitumen froth through dilution by paraffinic solvents. This review will compare asphaltene precipitation to dilution\textendash induced solvent shifting in aqueous systems via the ouzo effect. We attempt to highlight similarities and differences in the evolution of nanodroplets and effects from solvent mixing conditions and point out mutual experimental techniques and modeling approaches for the research of both asphaltene precipitation and nanodroplet formation.   The review will start from basic concepts in both asphaltene precipitation and the ouzo effect, then move on to the effects of solvent mixing on asphaltene precipitation in bulk. After that, we will introduce advances in microfluidic systems combined with cutting\textendash edge experimental and simulation tools for asphaltene precipitation induced under controlled mixing with solvents and a droplet formation. Such comparison will inspire more in\textendash depth understanding and control of asphaltene precipitation and the ouzo effect in aqueous systems. 
	
\end{abstract}


\end{frontmatter}

\cleardoublepage
\section{Introduction}
Asphaltene is the most complex and heaviest component in bitumen, consisting of polycyclic aromatic species with different molecular weights, polarity, and heteroatom content. Asphaltene is insoluble in paraffinic solvents and soluble in aromatic solvents \cite{taylor2001refractive}. Asphaltene precipitation is triggered by adding paraffinic solvents at the solvent/bitumen ratio (S/B ratio) above a critical value \cite{rao2013froth,xu2018}, a phenomenon of significance for industrial processes. For instance, solids and water are removed from bitumen froth in paraffinic froth treatment (PFT) through dilution by paraffinic solvents \cite{hristova2021bitumen}. For several years, numerous research has been done to recognize molecular or colloidal structures, solubility, and phase behaviour of asphaltene in different solvents and bitumen. Furthermore, asphaltene precipitation involves dynamic processes on multiple length scales (from nanometers to millimeters) and time scales (from seconds to days) However, dilution\textendash induced asphaltene precipitation is still one of the most significant and challenging topics due to the complexity of the physical and chemical properties of asphaltene. 

In parallel, dilution\textendash induced liquid\textendash liquid phase separation is omnipresent in many other technological or industrial processes. An example is the Ouzo effect in a ternary mixture of oil, ethanol, and water \cite{zhang2015formation}. When water is added to a clear Greek alcoholic drink Ouzo, anise oils become oversaturated due to dilution and form nanodroplets spontaneously in the cloudy mixture \cite{vitale2003liquid}. The same effect can also be seen when eucalyptus disinfectants and diluted with water \cite{zemb2016explain}. In liquid-liquid microextraction, oil microdroplets separated by dilution are the basic unit to concentrate and separate trace analytes from aqueous samples \cite{rezaee2006determination}. Gas molecules, lipids or polymers dissolved in a polar organic solvent all exhibit similar effects, forming bubbles or nanoparticles when diluted by a poor solvent. The process is also called as solvent exchange \cite{zhang2008nanobubbles}, nanoprecipitation \cite{schubert2011nanoprecipitation}, solvent displacement, solvent shifting \cite{aubry2009nanoprecipitation,lepeltier2014nanoprecipitation}, or flash precipitation.\\

\begin{table}
	\centering
	\caption{\large{Key existing review articles in the last two years.}}
	\label{literature}
	\renewcommand{\arraystretch}{0.6}
	\begin{tabular}{|p{5cm}|p{5cm}|p{0.8cm}|p{0.5cm}|p{0.5cm}}
		\hline
		Title & Subject of study & Year & \multicolumn{2}{l|}{Authors} \\ \hline
		Distributed   Properties of Asphaltene Nano-aggregates in Crude Oils: A Review & Molecular properties of   asphaltene, particle size distribution of nano-aggregates and their effect on   asphaltene precipitation. Molecular weight of nano-aggregates varies, and can be up to 4,0000 Da. The aggregates may be smaller than 100 $nm$.  & 2021 & \multicolumn{2}{l|}{ \citet{gray2021distributed}} \\ \hline
		Lab\textendash on\textendash a\textendash Chip   systems in asphaltene characterization: A review of recent advances & Research progress on asphaltene   precipitation by microfluidic devices. The confinement in microfluidics may have effects on the rheological properties of crude oil. 
		& 2021 & \multicolumn{2}{l|}{ \citet{mozaffari2021lab}} \\ \hline
		Asphaltene   precipitation and deposition: A critical review & Comparison of models of   asphaltene deposition to surfaces with experimental results. Asphaltenes with different molecular structures may exhibit different deposition models, and controversy on asphaltene structure hinders development of a unified asphaltene precipitation model.
		& 2021 & \multicolumn{2}{l|}{\citet{mohammed2021asphaltene}} \\ \hline
		A review on chemical sand production control techniques in oil reservoirs & Introduction and evaluation of methods for controlling and suppressing sand formation in oil reseviors & 2022 & \multicolumn{2}{l|}{\citet{saghandali2022review}} \\ \hline
		Crude   oil oxidation in an air injection based enhanced oil recovery process:   Chemical reaction mechanism and catalysis & Progress of air   injection for heavy oil recovery, summarize the problems faced at the current   stage and propose possible solutions & 2022 & \multicolumn{2}{l|}{\citet{yuan2022crude}} \\ \hline
		Perspectives on microfluidics for the study of asphaltenes in Upstream Hydrocarbon  Production: A Minireview &  Asphaltene deposition and and emulsion  stability in microfluidic channals. Small dimensions in microfluidic devices shortens analysis time, generating high-quality and reliable data for understanding of dynamic behavior. & 2022 & \multicolumn{2}{l|}{\citet{sharma2022perspectives}} \\ \hline
	\end{tabular}
\end{table}

Up to now, it has rarely been discussed that asphaltene precipitation shares several key features of dilution\textendash induced formation of microdomains (droplets or nanoparticles) in aqueous systems via the ouzo effect. The phase separation appears as microphase separation rather than bulk phase separation, which means that the phase separation forms uniformly distributed droplets or particles. The size and distribution of microdomains are determined not only by the concentration of the compositions but also by the temporal and spatial characteristics of the mixing process of the solvents \cite{zhang2015formation}. The addition of solvent greatly affects the size distribution, morphological characteristics, and physical and chemical properties of the formed microdomains. The formation and growth of microdomains are affected by several factors, including solution composition \cite{lu2015solvent}, mixing conditions of the solvent and the solution \cite{zhang2015formation}, and subsequent interactions with microdomains in proximity \cite{xu2017collective}.  The dynamics of droplet formation via the ouzo effect have seldom been compared to the  dynamics of asphaltene precipitation. The latest reviews on this subject are listed in Table\ref{literature}. In this critical review, for the first time, we will draw a comparison between solvent\textendash induced asphaltene precipitation and nanodroplet formation via the ouzo effect with emphasis on effects from mixing conditions. 


\section{Thermodynamics of solvent\textendash induced droplet formation and asphaltene precipitation}

\subsection{Spontaneous droplet formation via ouzo effect}
\begin{figure}
	\centering
	\includegraphics[width=0.7\textwidth]{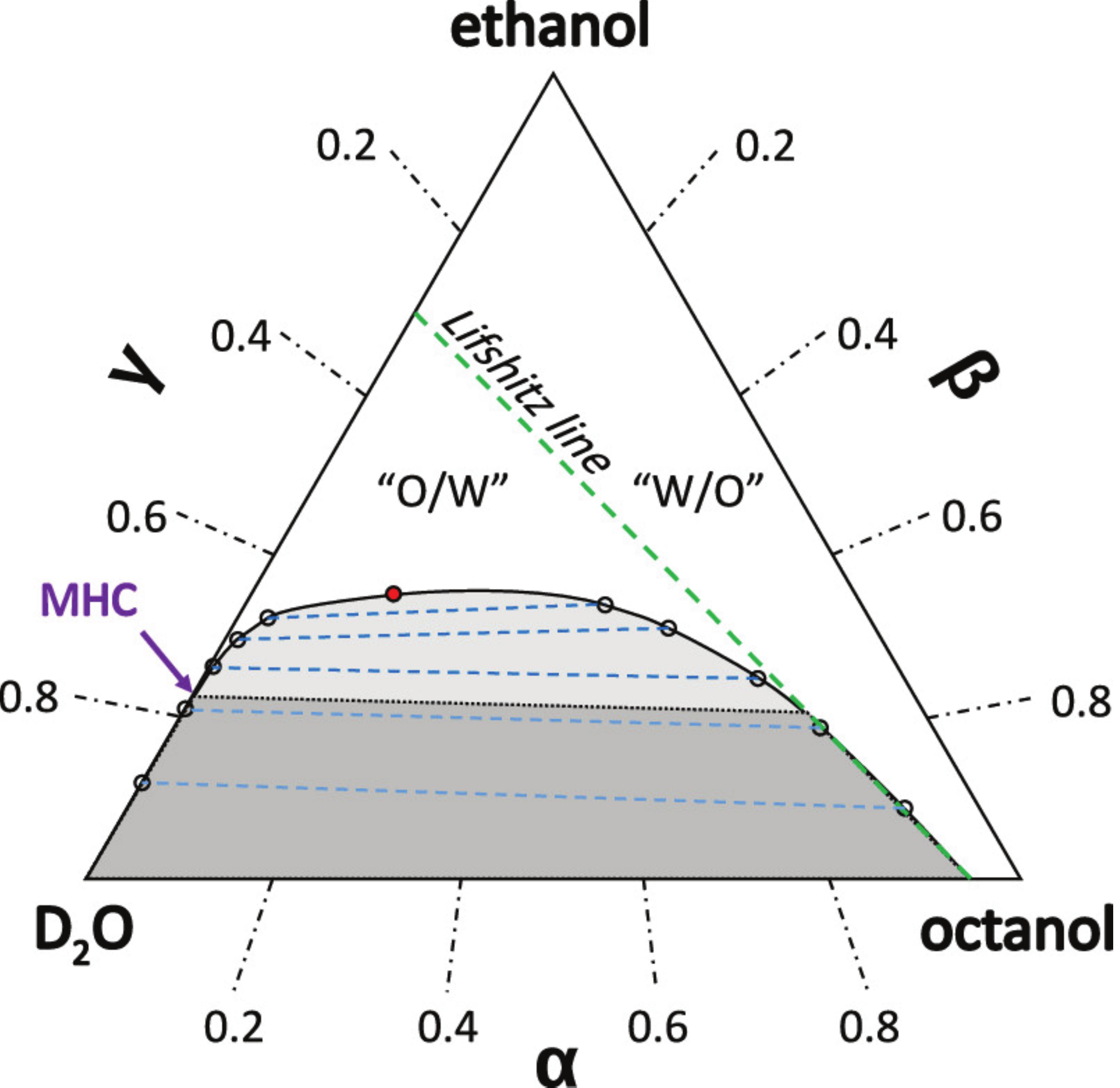}
	\renewcommand{\captionfont}{\linespread{1.6}\normalsize}
	\caption{Representative solubility phase diagram of the ternary system of oil, ethanol, and  heavy water. Where $\alpha$, $\beta$, and $\gamma$ are the mass fractions. Figure 1 Reproduced with permission from ref \cite{prevost2021spontaneous}. Copyright 2021 American Chemical Society.}
	\label{ouzo}
\end{figure}

Ouzo effect is the basis for liquid\textendash liquid microphase separation induced by dilution. Ouzo effect, also called spontaneous emulsification, is a milky oil\textendash in\textendash water emulsion obtained by adding water to an anise\textendash flavoured spirit. Such emulsion is formed by minimal mixing and is stable for a long time \cite{carteau2007probing,ganachaud2005nanoparticles,scholten2008life}. Recently, the Ouzo effect has been seen as a technique for the large\textendash scale generation of surfactant\textendash free emulsion without mechanical agitation \cite{qian2019surface}. A standard Ouzo effect system consists of three parts: oil, non\textendash solvent (i.e., water), and co\textendash solvent (i.e. ethanol). A typical phase diagram of the ternary system is shown in Figure \ref{ouzo} \cite{li2019controlled,lu2015solvent,wang2019formation}.  The spontaneous formation of microdomains takes place not only for liquid droplets in the case of the ouzo effect but also for polymer or nanoparticles, where the process is often referred to as nanoprecipitation \cite{schubert2011nanoprecipitation}. 

Oil is miscible with ethanol but immiscible with water. The ternary solution is a homogeneous solution with the composition above the binodal curve. With the addition of water, the oil solubility decreases, leading to oil oversaturation to form a new phase when the composition is below the binodal curve. The Ouzo region is a metastable region in which the Gibbs free energy is not a minimum, but there is a large kinetic barrier for phase separation, resulting in a stable emulsion\cite{vitale2003liquid}. In this process, the total volume of droplets generated is determined by the phase diagram, and the final volume of droplets increases with the oil concentration \cite{lepeltier2014nanoprecipitation}.

\begin{figure}
	\centering
	\includegraphics[width=0.8\textwidth]{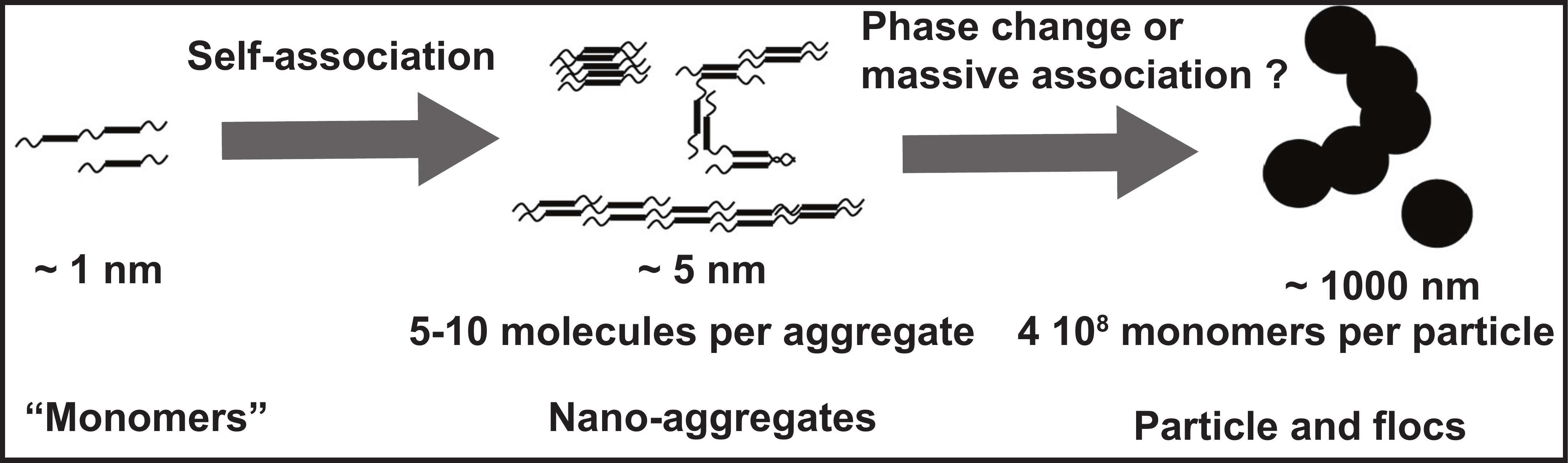}
	\renewcommand{\captionfont}{\linespread{1.6}\normalsize}
	\caption{Molecular, nano-aggregate, and precipitate length scales of asphaltene. Figure 2 Reproduced with permission from ref \cite{yarranton2013size}. Copyright 2013 American Chemical Society.}
	\label{mechanism}
\end{figure}

\subsection{Asphaltene precipitation induced by diluents}
As the phase separation process can be triggered by the addition of non-solvents, the physico and hydrodynamical aspects of dilution-induced asphaltene precipitation are similar to the droplet formation caused by ouzo effect. In the asphaltene precipitation process, as shown in Figure \ref{mechanism}, a few molecules first form nano\textendash aggregates through self\textendash association \cite{yarranton2013size}. Then, nano\textendash aggregates form particles or flocs revealed as phase separation.  The onset and quantity of asphaltene precipitation are related to the solvent type, S/B ratio, temperature, and pressure \cite{soleymanzadeh2019review,taylor2001refractive}.
Yen\textendash Mullins model proposed that there was a cluster state ($\sim$ 5 $nm$) between nano\textendash aggregates and phase separation states \cite{mullins2012advances}. The asphaltene aggregation process consists of $\pi$\textendash$\pi$ bonding between PAHs as well as other molecular interactions such as hydrogen bonding and acid\textendash base interactions \cite{schulze2015aggregation,gray2011supramolecular}. After the formation of the nano\textendash aggregates, steric repulsion restricts adding more molecules \cite{yarranton2013size}.

In aromatic\textendash based solutions, asphaltene colloids are in a favourable liquid environment, leading to the swelling and extended structures of asphaltene colloids. The asphaltene colloids repel each other due to the steric repulsion originating from the extended chains. In paraffinic\textendash based solvents, the liquid surrounding becomes less favourable for stretching of the asphaltene colloids, resulting in the reduction of the steric repulsion. The van der Waals attraction leads the aggregation of the asphaltene colloids to grow to large particles \cite{gray2011supramolecular,zahabi2010flocculation,wang2009colloidal,wang2010interaction}.

The influence of the type of the precipitant on the asphaltene precipitation is due to the difference in Hildebrand solubility parameters, which is different for different types of n\textendash alkanes \cite{barton2017crc}. The greater difference in the Hildebrand solubility parameters between solvent and asphaltene means the stronger attractive interaction between asphaltene particles \cite{wang2010interaction}. In addition to n\textendash alkanes, some other chemicals can also be used as a precipitant to induce asphaltene precipitation, such as carbon dioxide \cite{fakher2019asphaltene}. 

\citet{xu2018} attributed the effects of the S/B ratio and type of the precipitant on asphaltene precipitation to differences in the Hildebrand solubility parameter ($\delta$). As shown in Figure \ref{bulk_results}(a), the total asphaltene precipitation decreases with the increase of $\delta$ due to asphaltene becoming more soluble in the solvents.

\begin{figure}
	\centering
	\includegraphics[width=0.7\textwidth]{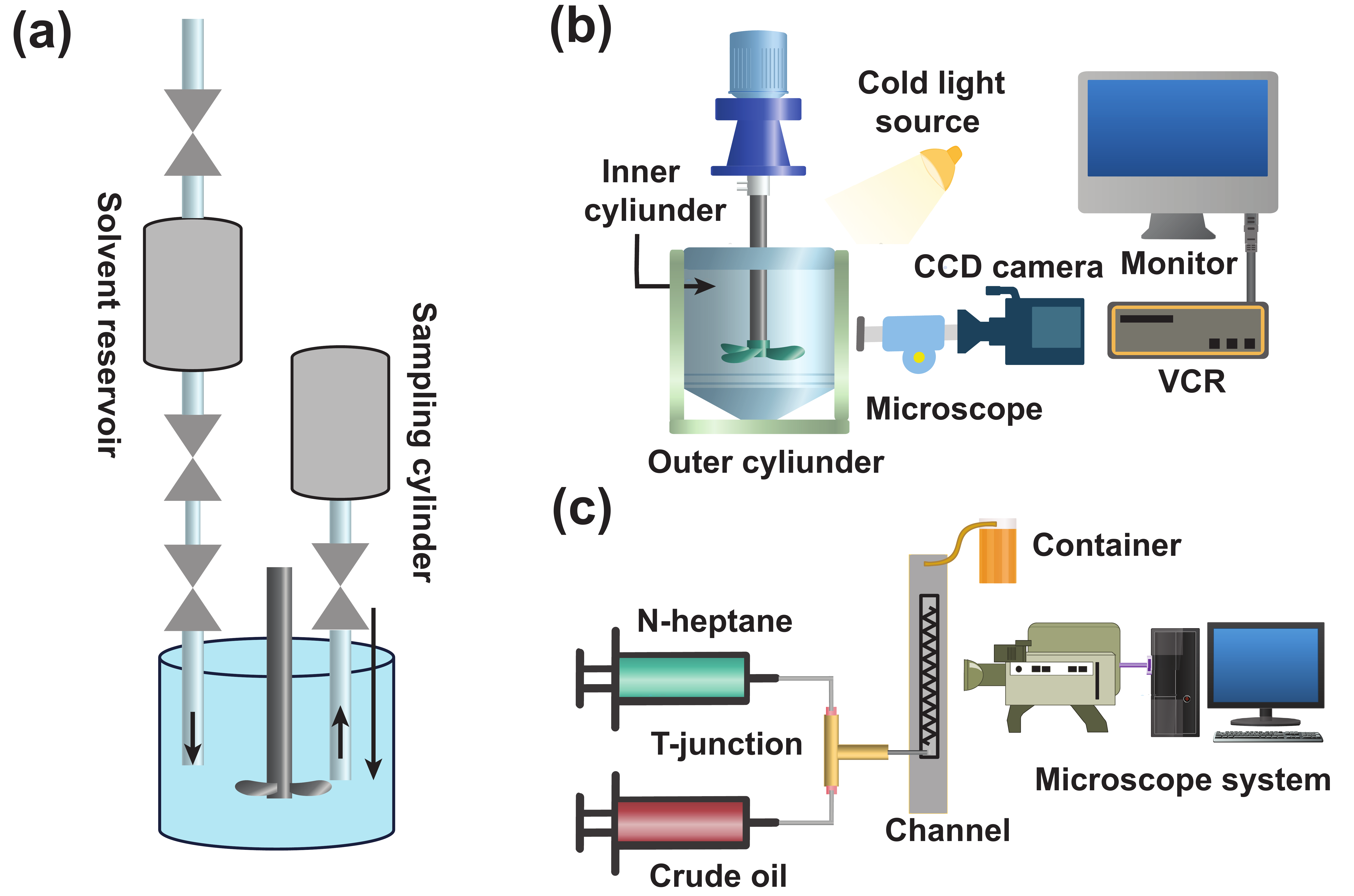}
	\renewcommand{\captionfont}{\linespread{1.6}\normalsize}
	\caption{Standard apparatus for asphaltene precipitation studies: (a) Bulk mixing system, (b) constant shear rate system, and (c) co\textendash injection system. Panel (a) reproduced with permission from ref \cite{xu2018}. Copyright 2017 American Chemical Society; Panel (b) reproduced with permission from ref \cite{rahmani2003characterization}. Copyright 2003 John Wiley \& Sons, Inc.; Panel (c) reproduced with permission from ref \cite{zhuang2016three}. Copyright 2016 Elsevier.}
	\label{setup}
\end{figure}
Solvent/bitumen ratio (S/B ratio) is the mass ratio of the paraffinic solvent to bitumen. Asphaltene precipitation happens when the S/B ratio is higher than the onset. But the onset measured by different methods may not be the same. \citet{tavakkoli2015indirect} found that the onset measured by indirect methods such as the near\textendash infrared method (NIR) is lower than that obtained from a direct method such as microscopy. This phenomenon is because different sizes of the smallest particles are observed by different methods. When the equipment was able to detect smaller particles, the measured S/B ratio would be smaller.

Several researchers found that the rate of asphaltene precipitation increases with the increase of the S/B ratio in terms of the yield and size of asphaltene \cite{li2018experimental,alboudwarej2003regular}. The interaction between asphaltene particles is stronger at a higher S/B ratio, resulting in higher rates of asphaltene precipitation and aggregation \cite{wang2010interaction}.  However, it is also reported that the change in the interaction between asphaltene particles also leads to a noteworthy change in the mechanism of asphaltene aggregation from diffusion\textendash limited aggregation (DLA) to reaction\textendash limited aggregation (RLA), particularly with the decrease in the S/B ratio \cite{balestrin2019direct,yudin1998mechanisms}. Two classical models were applied to describe the growth dynamics based on the mean diameter of the precipitates: diffusion\textendash limited aggregation (DLA) and reaction\textendash limited aggregation (RLA) \cite{yudin1998mechanisms,li2017experimental}. Although the mechanism of this transition was not fully understood, transitions from one model to another were observed at different S/B ratios, 
\cite{yudin1998mechanisms,li2017experimental,seifried2013kinetics}.

\section{Effects of mixing conditions on asphaltene precipitation}

Different from varying the characteristics of precipitates by mixture composition, the mixing dynamics in addition of the poor solvent may impact on the total amount and size distribution of the generated droplets or particles without changing the ratio of the antisolvent, due to the non-equilibrium nature before the homogeneous mixture is achieved. It is known that mixing conditions significantly impact the Ouzo effect from nucleation to the growth of the droplets. \citet{johnson2003mechanism} work reported that faster mixing resulted in a reduction in nanoparticle size induced by the ouzo effect. \citet{zhao2020particle} also reported that the particle size decreased with the increase of the speed of rotating by changing the agitator speed. \citet{beck2021solvent} work described when the ratio of solvent/non\textendash solvent increased, and the precipitation was more sufficient, finally leading to the generation of smaller nanoparticles. The effect of mixing conditions on the Ouzo effect is attributed to the timescale required for the system to reach homogeneous mixing, before which the local oversaturation varies with time. 

The effect of mixing dynamics on dilution\textendash induced liquid\textendash liquid phase separation has been extensively studied \cite{qian2019surface}. Many hydrodynamic factors of the mixing process affect the morphology of the resulting droplet, including flow rate \cite{zhang2015formation}, diffusion coefficient \cite{li2022growth}, viscosity \cite{meng2020viscosity}, and profile of the flow \cite{yu2015gravitational}. The mixing dynamics also affect the asphaltene precipitation, considering the similarity of these two processes.

\subsection{Shear rate in a mixing tank}

Previous studies on asphaltene precipitation mainly focused on the influence of thermodynamic factors, including S/B ratio \cite{yarranton2021regular}, type of the precipitant \cite{enayat2020development}, and temperature \cite{kuang2018investigation}. As represented in Figure \ref{setup}(a), a bulk mixing system was used. Results from the bulk mixture on asphaltene precipitation were studied as summarized in Table \ref{mixing summary}. 

\begin{table}[ht]
	\captionsetup{font=normalsize}
	\centering
	\caption{Asphatene precipitation in bulk by mixing.}
	\label{mixing summary}
	\begin{tabular}{|c|c|c|c|}
		\hline
		Mixing mechanism & Oil & Solvent & Reference \\ \hline
		\multirow{4}{*}{Agitator} & \begin{tabular}[c]{@{}c@{}}Venezuela \\ bitumen\end{tabular} & n\textendash Heptane & \cite{guo2017} \\ \cline{2-4} 
		& Crude oil & n\textendash Heptane & \cite{mohammadi2011inhibition} \\ \cline{2-4} 
		& \begin{tabular}[c]{@{}c@{}}McMurray \\ bitumen froth\end{tabular} & \begin{tabular}[c]{@{}c@{}}n\textendash Pentane, \\ isopentane, \\ neopentane\end{tabular} & \cite{xu2018} \\ \cline{2-4} 
		& \begin{tabular}[c]{@{}c@{}}Athabasca \\ bitumen froth\end{tabular} & \begin{tabular}[c]{@{}c@{}}n\textendash Pentane, \\ n\textendash Hexane, \\ n\textendash Heptane\end{tabular} & \cite{long2004structure} \\ \hline
		\multirow{2}{*}{Hand shaked} & Crude oil & n\textendash Heptane & \cite{haji2013unified} \\ \cline{2-4} 
		& \begin{tabular}[c]{@{}c@{}}Caoqiao \\ crude oil\end{tabular} & C5 \textendash  C12 & \cite{hu2001effect} \\ \hline
		\multirow{2}{*}{Pour into} & \begin{tabular}[c]{@{}c@{}}Karazhanbas \\ crude oil\end{tabular} & n\textendash Heptane & \cite{yudin1998mechanisms} \\ \cline{2-4} 
		& \begin{tabular}[c]{@{}c@{}}Maya crude oil, \\ Cold lake bitumen\end{tabular} & C5 \textendash  C16 & \cite{wiehe2005paradox} \\ \hline
		Sonicated & \begin{tabular}[c]{@{}c@{}}Cold lake \\ bitumen\end{tabular} & \begin{tabular}[c]{@{}c@{}}n\textendash Heptane, \\ n\textendash Pentane, \\ n\textendash Hexane, \\ n\textendash Octane\end{tabular} & \cite{alboudwarej2003regular} \\ \hline
	\end{tabular}
\end{table}

The mixing dynamics affects the movement of particles, which leads to the difference in particle collision frequency and in the formation of particles from nano-aggregates in Figure \ref{mechanism}. The collision frequency is determined by the movement of the particles, as shown in Equation (\ref{alpha}). Mechanical shearing of the mixture of crude oil and diluent significantly influences the settling properties of the aggregates and the size distribution of asphaltene precipitates. Large asphaltene\textendash solid aggregates with high density formed when solvent and bitumen froth was mixed at a higher speed for a long time \cite{zawala2012settling}. 
\citet{rahmani2005fractal} controlled the shear rate in a bulk system by rotating the rate of a cylinder, as shown in Figure \ref{setup}(b). Under the action of shear rate, asphaltene precipitation is an equilibrium between aggregation and fragmentation processes. The shear rate intensified the movement of asphaltene particles, enhancing the collision frequency \cite{rahmani2003characterization}. 

The difference in the settling rate of asphaltene particles also induces the collision of asphaltene particles. As a result, three factors contribute to the collision frequency of the system in Figure \ref{setup}(b), including Brownian motion, shear rate, and differential settling of asphaltene particles \cite{rahmani2004evolution}. 

\begin{subequations}
	\label{alpha}
	\begin{align}
		Brownian: \alpha_{i,j} &= \frac{2}{3} \frac{RT}{\mu} \frac{(d_i + d_j)^2}{d_i d_j}\\
		Shear: \alpha_{i,j} &= \frac{G}{6} (d_i + d_j)^3\\
		Settling: \alpha_{i,j} &= \frac{\pi g}{72 \mu} (d_i + d_j)^2 |\Delta \rho_i d_i^2 - \Delta \rho_j d_j^2|
	\end{align}
\end{subequations}
where $\alpha_{i,j}$ is collision frequency, $R$ is ideal gas constant, $\mu$ is viscosity, $T$ is temperature, $G$ is shear stress, $d_i$ and $d_j$ are the diameter of particle $i$ and $j$, respectively, $\rho_i$ and $\rho_j$ are the density of particle $i$ and $j$, respectively.

The pressure imbalance between the two sides of the asphaltene particles caused by the shear rate also causes the fragmentation \cite{torkaman2018influence}. Fragmentation results in the change of large particles into small particles, which limits the growth of particle size \cite{nguyen2021effect}. Considering collision and fragmentation, the growth of asphaltene particles can be calculated by the Smoluchowski aggregation model \cite{rahmani2004evolution}:

\begin{equation}
	\label{Sm_2}
	\frac{dn_k}{dt}=\frac{1}{2} \sum_{i+j=k} \alpha_{i,j} \beta_{i,j} n_i n_j - n_k \sum_{i \ge 1} \alpha_{i,k} \beta_{i,j} n_i - B_k n_k + \sum_{j=k+1}^{n_{max}} \gamma_{k,j} B_j n_j
\end{equation}
where $t$ is time, $n_k$ is the number density of particles of size $k$, $\beta_{i,j}$ is the collision efficiency, which depends on the interaction between asphaltene particles in the solvents, $B_k$ and $B_j$ are the fragmentation rate of particles $k$ and $j$, $n_{max}$ is the largest particle size form fragments of size $k$ by breakage, $\gamma_{k,j}$ is the volume fraction of the fragments of size $i$ originating from size $j$ particles. The fragmentation rate ($B_k$) is a function of particle volume ($V_k$) \cite{pandya1983floc,solaimany2008dynamic}:

\begin{equation}
	\label{Bk}
	B_k = bV_k^{1/3}
\end{equation}
where b is a correlation parameter \cite{rahmani2004evolution}. 

The existence of shear rate enhances aggregation and fragmentation of asphaltene particles simultaneously \cite{igder2018control}. At the beginning of mixing of asphaltene solution and the paraffinic solvents, the aggregation dominates the process, leading to the size increase of asphaltene particles until a critical value. The fragmentation then dominates the process, resulting in the size decreases. At the final state, aggregation and fragmentation reach an equilibrium, and the size of asphaltene particles remains stable \cite{rahmani2003characterization}. With the increase in shear rate, the rate of aggregation of asphaltene particles becomes faster, but the rate of fragmentation also increases. Therefore, the size of asphaltene particles at the final state decreases with the increase of shear rate, as shown in Figure \ref{bulk_results}(b) \cite{rahmani2004evolution}. The precipitation dynamics of asphaltene particles are the same at all points in the system, as shown in Figure \ref{bulk_results}(c). Therefore, the local composition does not affect asphaltene precipitation in the constant shear rate system.

\subsection{Mixing in a narrow channel}

Microfluidics has become a powerful approach for the reproducible synthesis of nanomaterials by nanoprecipitation. 
Microfluidics offers the possibility to produce nanoparticles in a continuous flow, facilitating high\textendash quality synthesis of large amounts. In addition, it could be further scaled\textendash up through parallel fabrication.  
The dynamics of phase separation can be achieved by co\textendash injection of a mixture of oil, a good solvent and a poor solvent into a narrow channel. \citet{chen2022} found that for the channel of cross\textendash shaped, an increase in flow rate leads to an increase in nanoparticle size. However, for the channel of multi\textendash lamination, the increase in flow velocity leads to a decrease in nanoparticle size. In the split\textendash recombine channel, the flow rate has almost no effect on the particle size. Using COMSOL simulation, \citet{zhao2020particle} found that the influence of flow velocity change on the dynamics of precipitation was due to the difference in solution mixing. Faster mixing results in smaller droplets or particles. For channels with different dimensions, the effect of flow rate on solution mixing is different. Therefore, the influence of flow velocity change on precipitation kinetics in different channels is different \cite{abdelkarim2021microchannel}. To show the importance of controlling the mixing conditions for polymer nanoparticles loaded with a fluorescence dye and identify the optimal microfluidic setup for the production of polymer nanoparticles for biomedical applications, \citet{chen2022} compared  the size and fluorescence properties of synthesized nanoparticles by using 5 different microfluidic mixers. The flow rate significantly impacted the nanoparticles' size and fluorescence properties. Impact jet mixers strongly decrease the size of nanoparticles due to strong enhancement in the mixing speed. The mixing speed also has an effect on the enhancement of the fluorescence properties of the nanoparticles.

Co\textendash injection is a method for the study of asphaltene precipitation under different flow conditions. The mechanism is shown in Figure \ref{setup}(c). Asphaltene solution and the precipitant are injected from two syringes into a microcapillary. The asphaltene precipitation happens when mixing the asphaltene solution and the precipitant is well enough. The flow rate and chemical composition in the two syringes can be controlled. The microcapillary tube can be changed to substrates with different types of patterns to investigate the structure effect. Some typical studies are summarized in Table \ref{controlled mixing}.

\begin{table}[!ht]
	\centering
	\captionsetup{font=normalsize}
	\caption{Co\textendash injection for mixing to induce asphaltene precipitation by n\textendash heptane.}
	\label{controlled mixing}
	\begin{tabular}{|c|c|c|}
		\hline
		Mixing device & Oil & Reference \\ \hline
		\multirow{4}{*}{\begin{tabular}[c]{@{}c@{}}Co\textendash injection into \\ a microcapillary\end{tabular}} & C7 asphaltene in toluene & \citet{boek2008deposition} \\ \cline{2-3} 
		& South America heavy oil & \citet{lawal2012experimental} \\ \cline{2-3} 
		& Venezuela crude oil & \citet{li2018experimental} \\ \cline{2-3} 
		& Crude oil & \begin{tabular}[c]{@{}c@{}} \citet{boek2010multi}, \citet{zhuang2016three,zhuang2018experimental} \end{tabular} \\ \hline
		\multirow{4}{*}{\begin{tabular}[c]{@{}c@{}}Co\textendash injection into a \\ microchannel with micropatterns\end{tabular}} & \multirow{2}{*}{C7 asphaltene in toluene} & \citet{lin2019microfluidic} \\ \cline{3-3} 
		&  & \citet{hu2014microfluidic} \\ \cline{2-3} 
		& \multirow{2}{*}{C5 asphaltene in toluene} & \citet{lin2016examining} \\ \cline{3-3} 
		&  & \citet{lin2017characterizing}\\ \hline
		\begin{tabular}[c]{@{}c@{}}n\textendash Heptane flux mixes \\ with model oil in cavity\end{tabular} & Asphaltene in benzene & \citet{shalygin2019spectroscopic} \\ \hline
	\end{tabular}
\end{table}

\begin{figure}[!ht]
	\centering
	\includegraphics[width=0.75\textwidth]{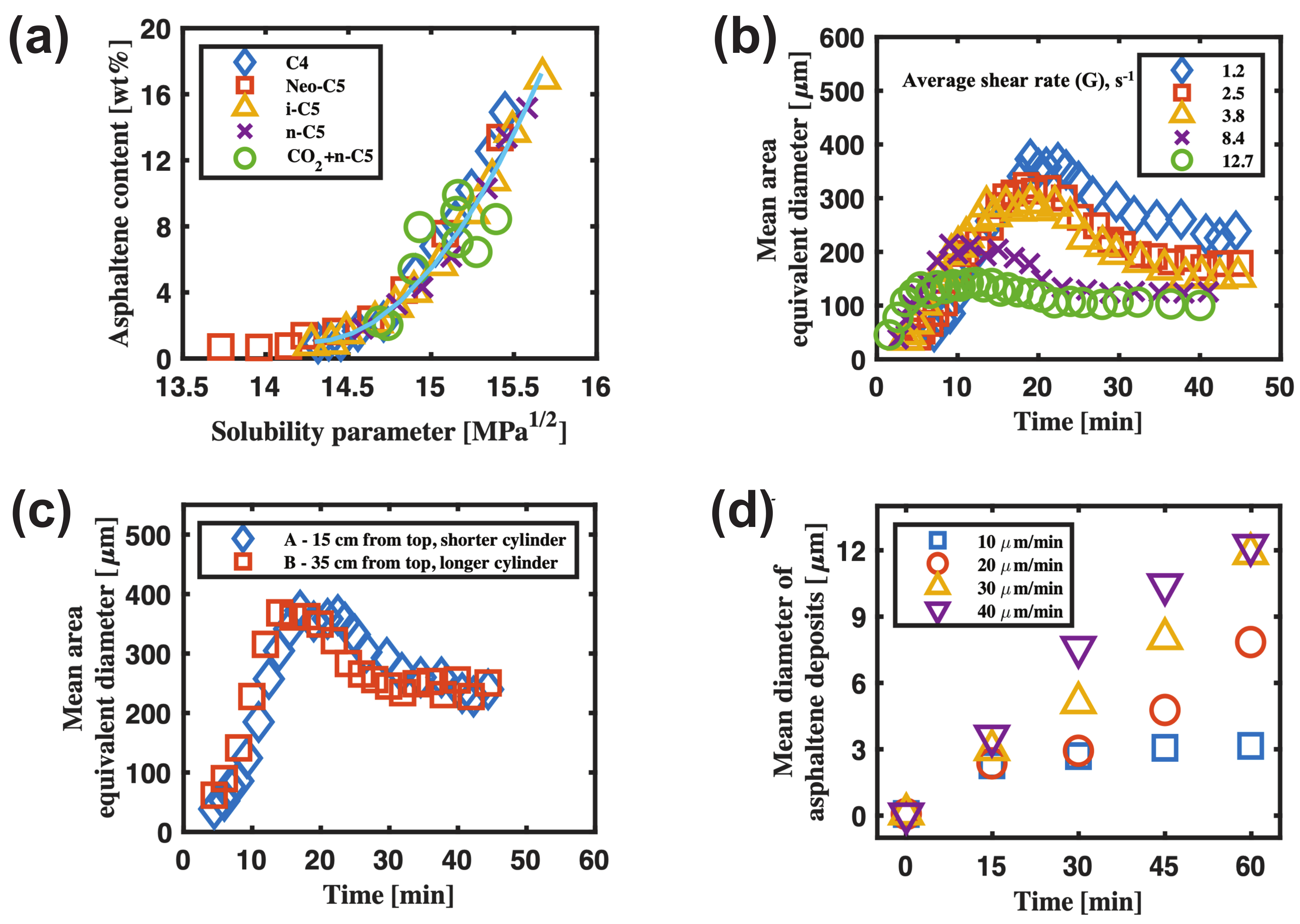}
	\renewcommand{\captionfont}{\linespread{1.6}\normalsize}
	\caption{(a) Relationship between asphaltene residues in bitumen with Hildebrand Solubility parameter in a bulk mixing system. (b) Growth kinetics of asphaltene particles under different shear rates and (c) comparison between different locations in a constant shear rate system. (d) Growth kinetics comparison of asphaltene induced by different flow rates in a co\textendash injection system. Panel (a) reproduced with permission from ref \cite{xu2018}. Copyright 2018 American Chemical Society;  Panels (b)  and (c) are reproduced with permission from ref \cite{rahmani2003characterization}. Copyright 2003 John Wiley \& Sons, Inc.; Panel (d) reproduced with permission from ref \cite{li2018experimental}. Copyright 2018 Elsevier.}
	\label{bulk_results}
\end{figure}

\citet{boek2008deposition,boek2010multi} measured the conductivity change of the microcapillary in the co\textendash injection process. The more conductivity decreases for a given time means, the higher rate of asphaltene deposition. They found that the increase in the injection rate of the solutions accelerates the deposition rate of asphaltene initially. But the microcapillary is not blocked by the asphaltene deposition at a higher flow rate (10 $\mu L/min$) because the deposits are removed by the flow. Using the same setup, \cite{zhuang2016three, zhuang2018experimental} found that the thickness of the asphaltene deposition layer increased with the flow rate from 10 $\mu L/min$ to 40 $\mu L/min$ and then decreased from 40 $\mu L/min$ to 100$\mu L/min$. The results were verified by \citet{zhuang2018experimental} using a 3D microscope. Later, by using the microscope to visualize the asphaltene particles, \citet{li2018experimental} found that the size of the asphaltene particles increases with the increase of the flow rate of injection of the solutions, as shown in Figure \ref{bulk_results}(d). 

\citet{guo2017} followed asphaltene aggregation and deposition on a microcapillary tube for the purpose of mimicking the situation in a porous medium. A faster flow rate was found to lead to larger asphaltene precipitates on the wall of the capillary tube \cite{li2018experimental}. But \citet{lawal2012experimental} reported that the deposition of asphaltene was independent of flow rate and was reasonably uniform along the capillary tube, in disagreement with earlier observation. The above\textendash reported effects from mixing after the solvent addition were attributed to interruption of the fractal structures of asphaltene precipitates or aggregate formation. There may also be effects from precipitate\textendash flow interactions or interactions among precipitates in the mixing flow. The deposition on the microcapillary tube may be influenced by both asphaltene phase separation and different precipitates\textendash wall under the shear along the tube. Furthermore, it is unknown how water drops influence the onset of asphaltene precipitation and their association with precipitates under mixing processes.

\begin{figure}[!ht]
	\centering
	\includegraphics[width=0.75\textwidth]{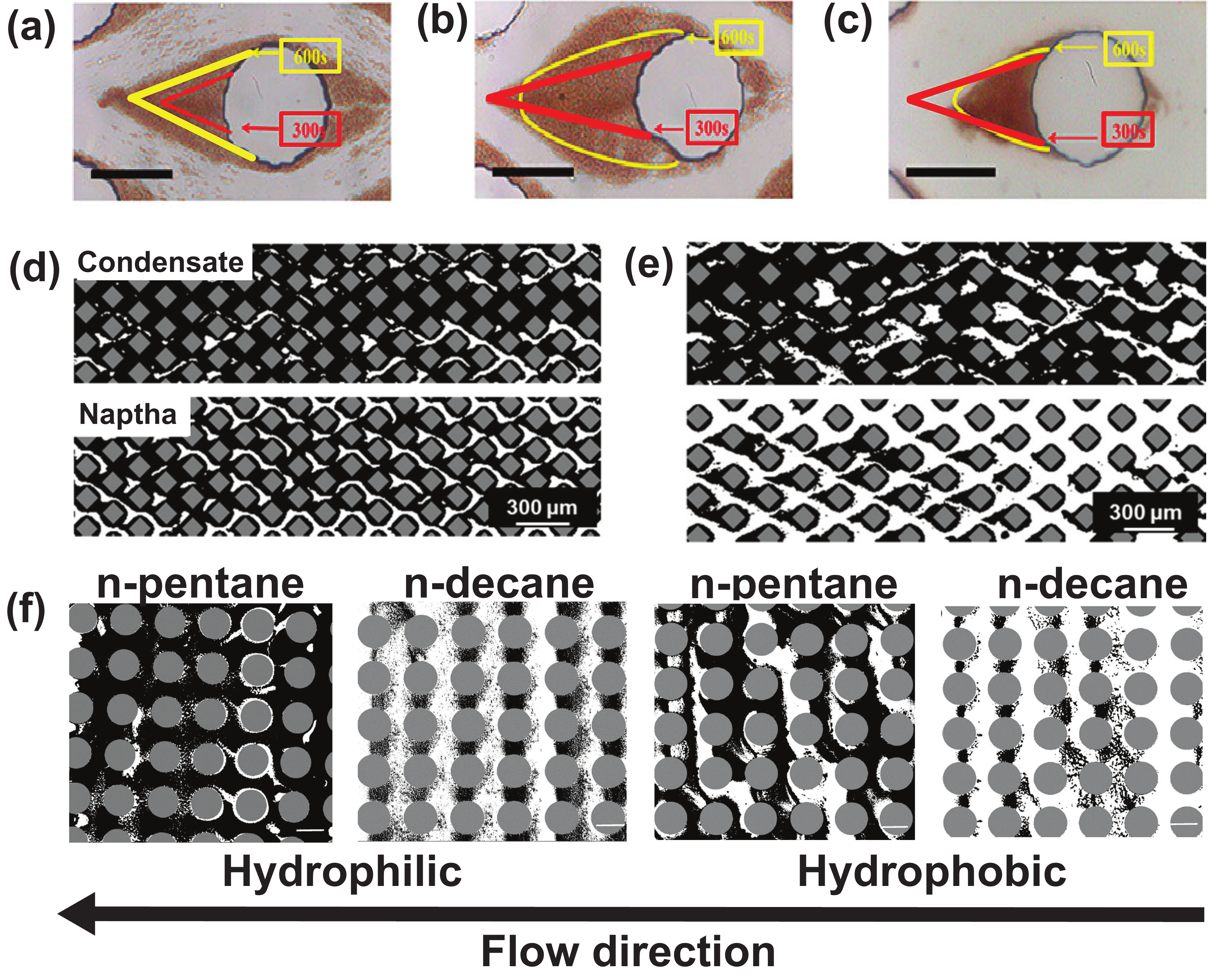}
	\renewcommand{\captionfont}{\linespread{1.6}\normalsize}
	\caption{Asphaltene deposition growth in different dispersants: (a) without dispersants, (b) with p\textendash octylphenol, and (c) with iso\textendash dodecylphenol. Deposition profiles of asphaltene in condensate and naphtha for (d) 50 $\mu m$ and (e) 100 $\mu m$ pore throat. (f) Deposition profiles of asphaltene in hydrophilic and hydrophobic surfaces. Panels (a)-(c) reproduced with permission from ref \cite{lin2017characterizing}. Copyright 2018 American Chemical Society; Panels (d)-(e) reproduced with permission from ref \cite{qi2018asphaltene}. Copyright 2017 American Chemical Society; Panel (f) reproduced with permission from ref \cite{keshmiri2019microfluidic}. Copyright 2018 Elsevier.}
	\label{microdevice}
\end{figure}
\citet{hu2014microfluidic} changed the microcapillary into a channel with porous structures. They found that the pores were easier to block when the solutions' injection rate was low. \citet{lin2016examining,lin2019microfluidic} found that the asphaltene deposits tended to grow against the flow direction rather than laterally because a high shear rate limited the growth of asphaltene, regardless of whether using a dispersant (Figure \ref{microdevice}(a)). The smaller asphaltene particles were easier to resist the shear from the flow \cite{lin2017characterizing}. And \citet{lin2021pore} found that asphaltene preferred to precipitate in the large pores rather than small pores. \citet{qi2018asphaltene} found that small pores were easier to block than large ones, as shown in Figure \ref{microdevice}(b). \citet{keshmiri2019microfluidic} changed the hydrophobicity of the surface and found that deposition of asphaltene on a hydrophilic surface is faster (Figure \ref{microdevice}(c)), which is caused by the higher interaction potential between asphaltene and hydrophilic surface. 

\citet{shalygin2019spectroscopic} found that the different flow affects the mixing and ultimately changes the local chemical composition so that the asphaltene precipitation in different positions in the substrate may be different. \citet{mokhtari2022asphaltene} also found that the higher local concentration of asphaltene at the water\textendash oil interface can accelerate the aggregation rate of asphaltene particles, ultimately enhancing asphaltene precipitation. Therefore, mixing dynamics not only affect particle movement and change particle collision frequency but also affect the local chemical composition and change the collision efficiency of asphaltene particles.

\section{Droplet formation and asphaltene precipitation from diffusive\textendash dominated mixing}
\subsection{Liquid\textendash liquid separation via ouzo effect in confined spaces}

Diffusive-dominated mixing is a well-controlled method to study the influence of mixing parameters on phase separation process. The experiment is more reproducible and the process is easier to be described by modelling. A series of publications from \citet{lu2017universal} reported that in ternary liquid\textendash liquid systems, the oversaturation level and growth duration of the new phase are mediated by P\'{e}clet number (Equation \ref{Pe}) of the flow, geometrical features of the fluid chamber, and the solution composition. A faster laminar flow of the solvent leads to larger droplets from the phase separation. The complex effects from solvent diffusion and mixing dynamics were decoupled for microphase separation in a confined 2\textendash dimensional channel.

\begin{figure}[!ht]
	\centering
	\includegraphics[width=0.90\textwidth]{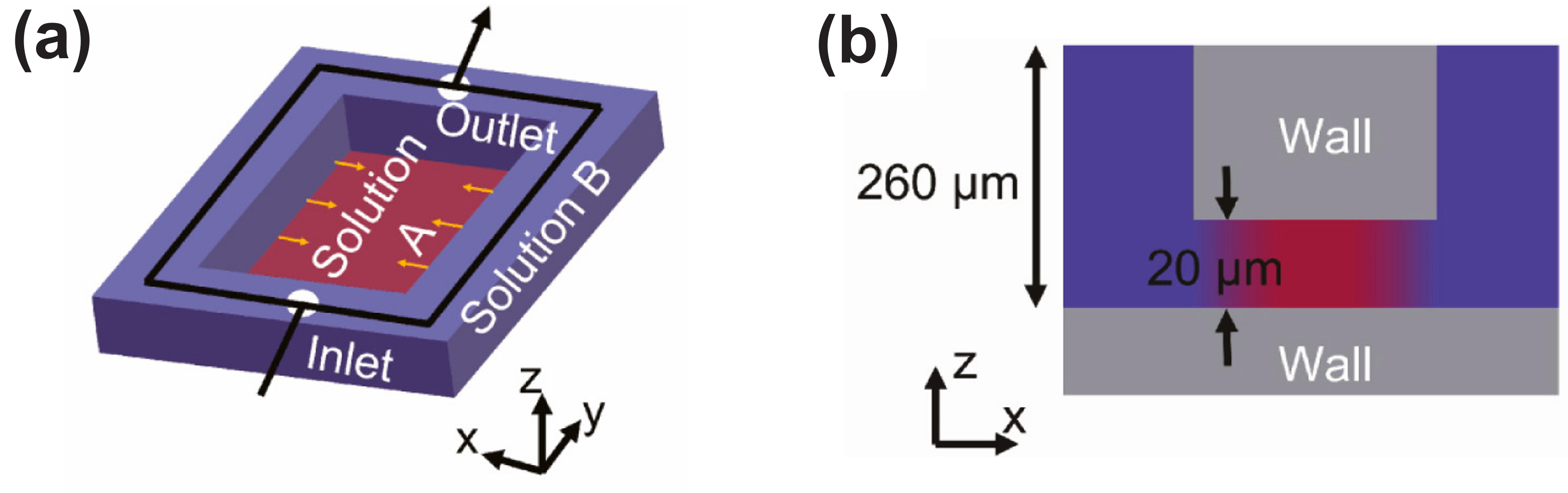}
	\renewcommand{\captionfont}{\linespread{1.6}\normalsize}
	\caption{(a) Sketch of the quasi\textendash 2D microchamber used in this study. The black arrow indicates the flow direction of the poor solvent. Yellow arrows indicate the diffusion direction of the poor solvent. (b) Side view of the cross\textendash section of the microchamber. Panels (a) and (b) are reproduced with permission from ref \cite{meng2022size}. Copyright 2022 Elsevier.}
	\label{chamber}
\end{figure}
\begin{figure}[!ht]
	\centering
	\includegraphics[width=0.7\textwidth]{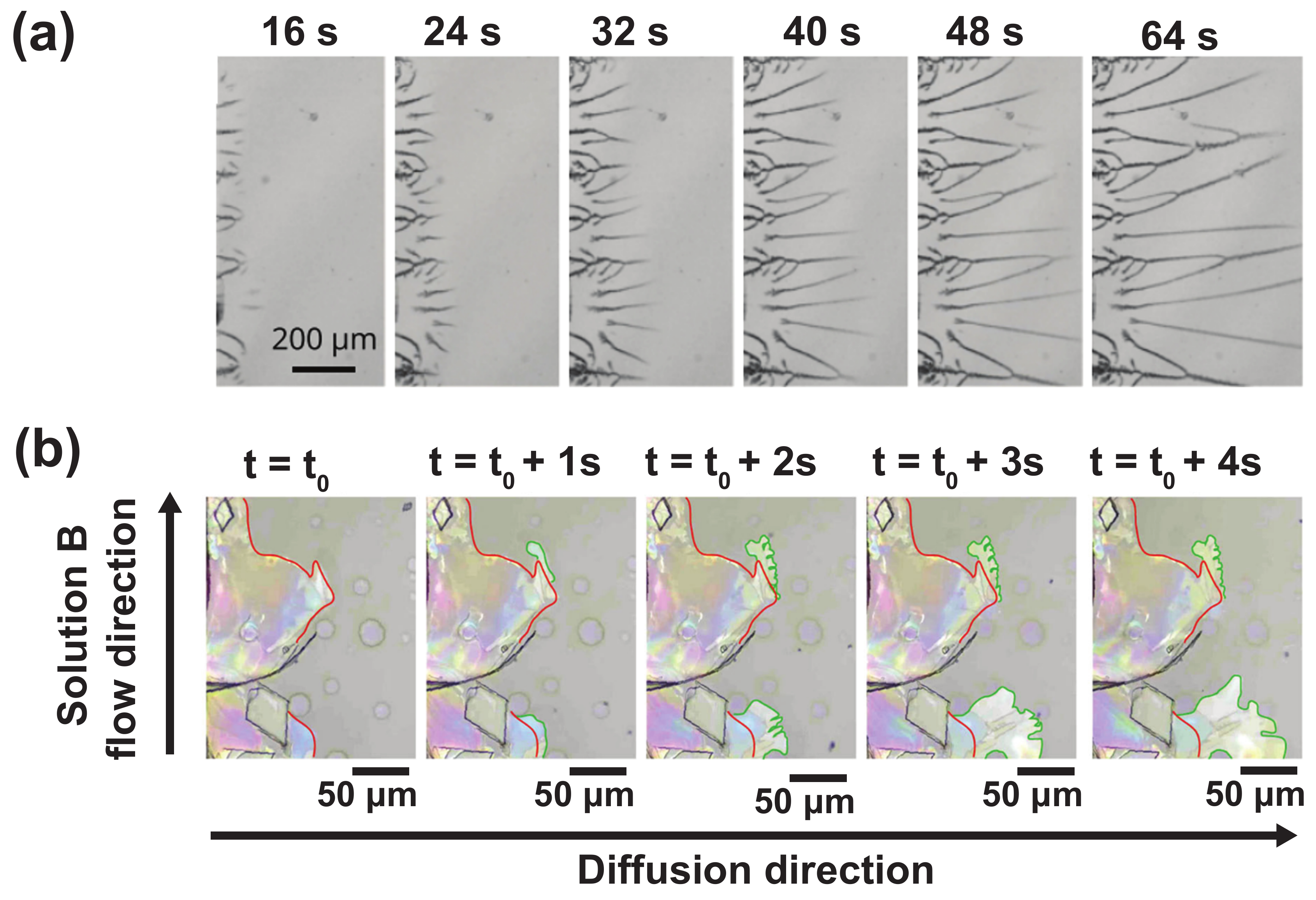}
	\renewcommand{\captionfont}{\linespread{1.6}\normalsize}
	\caption{(a) Growth of branch structure of precipitations and (b) crystallization in the quasi\textendash 2D chamber. Panel (a) reproduced with permission from ref \cite{lu2017universal}. Copyright 2017 National Academy of Sciences; Panel (b) reproduced with permission from ref \cite{meng2021microfluidic}. Copyright 2021 Springer Nature.}
	\label{quasi-2D}
\end{figure}

As sketched in Figure \ref{chamber}, the horizontal rectangular flow chamber was constructed by assembling two glass plates sealed with a spacer. The distance from the top plate to the bottom surface is very close (i.e., 20 $\mu m$) to avoid convective flow. The main flow chamber is flanked by two deep side channels (i.e., 260 $\mu m$), as indicated by the two rectangles in the sketch \cite{meng2021microfluidic}. Inside the quasi\textendash 2D fluid chamber, the concentration of the solvent was known both space\textendash wise and time\textendash wise \cite{lu2017universal}. 
In the study of the liquid\textendash liquid separation, diffusive\textendash dominated mixing causes the branch\textendash like the structure of droplets, as shown in Figure \cite{lu2017universal}, as shown in Figure \ref{quasi-2D}(a). \citet{arends2021enhanced,arends2021fast} found a jetting phenomenon of the droplets, and the speed of the jetting was related to the composition of the solution. \citet{meng2021microfluidic} successfully applied the proposed quasi\textendash 2D chamber to the study of liquid\textendash solid separation systems such as oiling\textendash out crystallization, as shown in Figure \ref{quasi-2D}(b).

\subsection{Visualization of asphaltene precipitation from diffusive-dominated mixing}

\begin{figure}[!ht]
	\centering
	\includegraphics[width=1\textwidth]{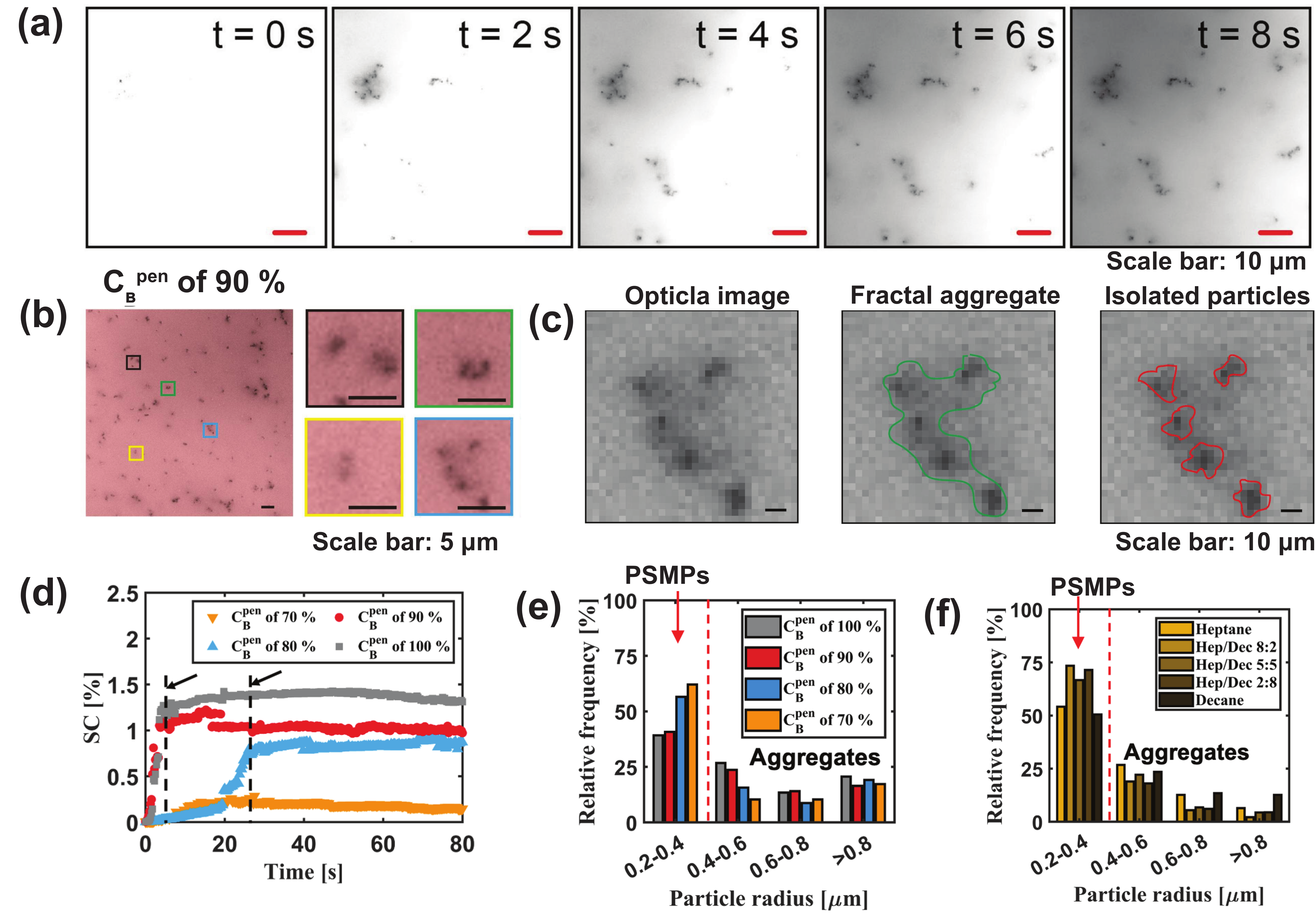}
	\renewcommand{\captionfont}{\linespread{1.6}\normalsize}
	\caption{(a) Snapshots of the growth dynamics of the asphaltene particles in the quasi\textendash 2D chamber. (b) Morphology of asphaltene particles at the final state. (c) Image analysis to separate the aggregates into isolated particles. (d) Growth dynamics of the surface coverage of asphaltene particles in the quasi\textendash 2D chamber. Size distribution of the asphaltene particles precipitated by (e) n\textendash pentane and (f) a mixture of n\textendash heptane and n\textendash decane.  Panels (a)-(e) are reproduced with permission from ref \cite{meng2021primary}. Copyright 2021 Elsevier;  Panel (f) reproduced with permission from ref \cite{meng2022size}.  Copyright 2022 Elsevier.}
	\label{PSMPs}
\end{figure}

Inspired by the phenomenon in the quasi\textendash 2D chamber, \citet{meng2021microfluidic} used the quasi\textendash 2D chamber to study the precipitation of asphaltene caused by dilution. Unlike the traditional bulk system, \cite{lim2018effect,duran2018nature}, asphaltene precipitation in the diffusive\textendash dominated mixture system is not a continuous process. As shown in Figure \ref{PSMPs}(a), for a specific position in the main channel, the solution composition gradually changed from pure asphaltene solution to pure n\textendash pentane, and asphaltene precipitation began at the moment that the n\textendash pentane concentration was higher than that of onset until asphaltene solution was totally displaced by n\textendash pentane. This process freezes the process of the early stage of asphaltene precipitation.

Direct visualization of the asphaltene precipitates from an early stage is challenging, including size distribution and structural characteristics, mainly because of the high optical density of the mixture of asphaltene components \cite{struchkov2019laboratory}. Confocal laser-scanning microscopy in a transmission mode was recently used as high-spatial-resolution measurements \cite{seifried2013kinetics}. The size distribution of asphaltene particles in different solvents was observed versus time when a very thin film of the mixture was confined between two plates \cite{seifried2013kinetics}. Confocal microscopy and image processing techniques were used to study the yield and size distribution of asphaltene particles after phase separation \cite{castillo2004study,hung2005kinetics}. In the above-mentioned optical measurements, the concentration of asphaltene must remain very low ($<$ 0.2 $mg/mL$) to allow enough intensity of the light passing through the samples \cite{taylor2001refractive}. Therefore, the measurements were either much lower than the concentration regimes in bitumen or the entire early stage dynamics of the precipitation were missed due to the high solvent ratio above the critical ratio.

\begin{figure}
	\centering
	\includegraphics[width=0.65\textwidth]{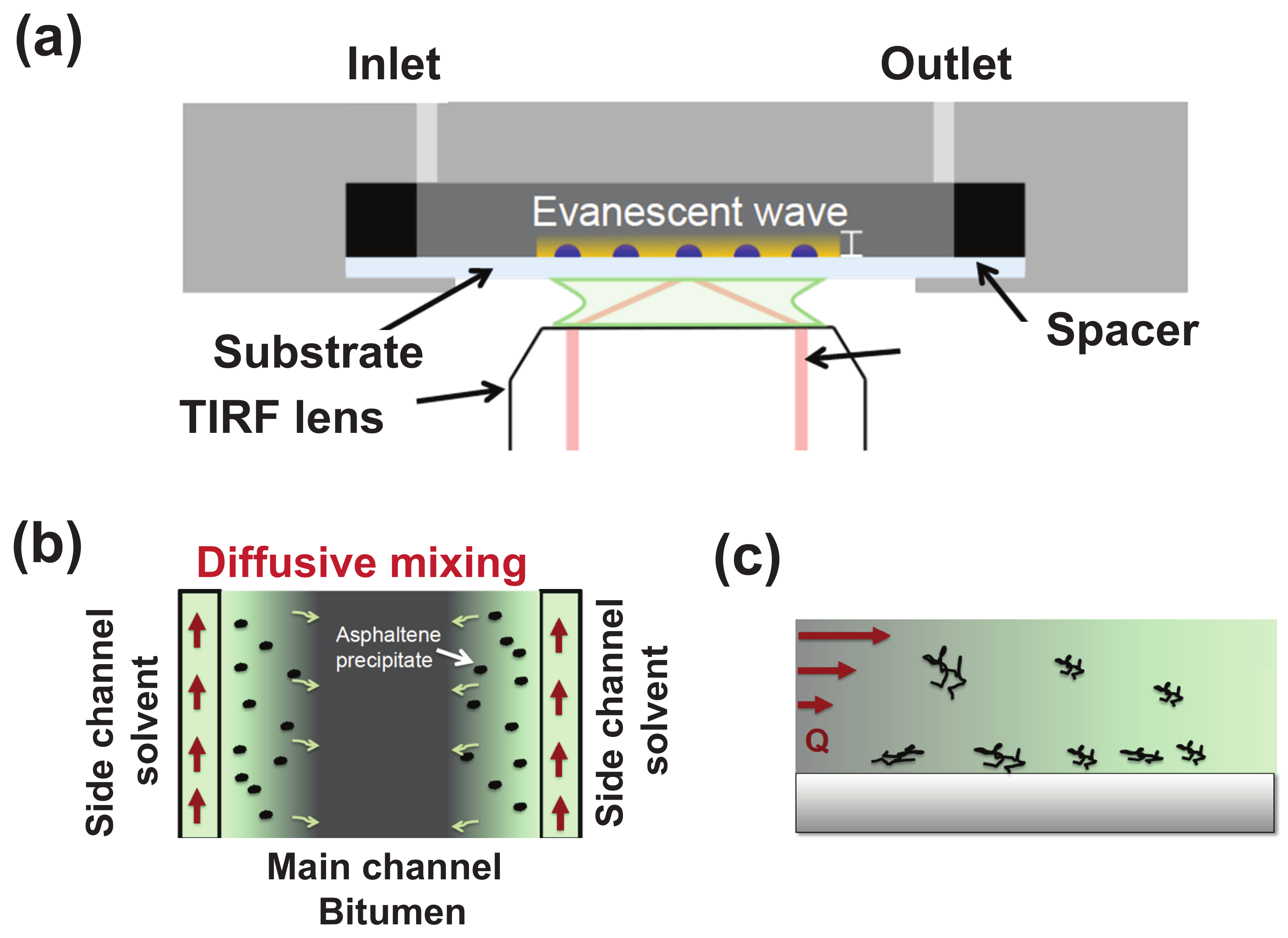}
	\renewcommand{\captionfont}{\linespread{1.6}\normalsize}
	\caption{(a) Sketch of the total internal reflection fluorescence microscopic system (TIRF) with the designed microfluidic chamber. (b) A quasi\textendash 2D channel for diffusion driven mixing. The channel has two deep\textendash side subchannels (in two rectangular boxes) and a very narrow main channel. (c) Sketch of the substrates for asphaltene precipitation on a planar surface \cite{meng2021microfluidic,zhang2015formation}. Panel (a) reproduced with permission from ref \cite{dyett2018growth}. Copyright 2018 The Royal Society of Chemistry. }
	\label{TIRF}
\end{figure}

Total internal reflection fluorescence microscopic system (TIRF) in Figure \ref{TIRF}(a) is a powerful technique to overcome experimental difficulties from asphaltene while providing high temporal and spatial resolutions \cite{guo2018single}. In the configuration of TIRF, the light propagates within the substrate by total internal reflection and generates an evanescent field from the substrate \cite{guo2018single}. The selective excitation and visualization of samples in TIRF are confined in the evanescent field generated from the total internal reflection. The thickness of the evanescent field is approximately 200 $nm$ adjacent to a substrate, depending on the incident angle and refractive index of the media \cite{niederauer2018direct}. Compared with a transmission or reflection mode of the conventional optical microscope, the most desirable feature of the TIRF configuration is that the high scattering background from the medium far from the substrate can be largely neglected \cite{dyett2018growth}. Therefore, the difficulty of the high optical density will be largely overcome. The unique capacity of TIRF allows for tracking the microphase separation during solvent mixing \cite{axelrod2021light}.

There are interesting insights obtained from combining diffusive mixing with TRIF measurements.  At the early stage of asphaltene precipitation, two types of asphaltene particles are formed, including individually existing particles and fractal aggregates, as shown in Figure \ref{PSMPs}(b). With image processing, the fractal aggregates can be separated into isolated particles (Figure \ref{PSMPs}(c)). The size of the isolated particles is comparable with the individually existing particles, and the size is always between 0.2 and 0.4 $\mu m$ (Figure \ref{PSMPs}(d)). Therefore, the particles from 0.2 to 0.4 $\mu m$ are defined as primary submicron particles (PSMPs), which means they are the basic blocking units for fractal aggregates in the asphaltene precipitation process. PSMPs are always formed in the studies of \citet{meng2021primary,meng2022size}, including different solvent concentrations and mixing conditions, and whether nonylphenol is used as inhibitor, as summarized in Table \ref{PSMPs_conditions}. 

\begin{table}[ht]
	\captionsetup{font=normalsize}
	\centering
	\caption{Summary of conditions to form PSMPs.}
	\label{PSMPs_conditions}
	
	\begin{tabular}{|c|c|c|c|}
		\hline
		Solution A                                                                                & Solution B       & Mixing conditions                                                                & Inhibitor            \\ \hline
		\multirow{21}{*}{\begin{tabular}[c]{@{}c@{}}17 g/L asphaltene\\  in toluene\end{tabular}} & 70 \% n\textendash pentane  & \multirow{20}{*}{\begin{tabular}[c]{@{}c@{}}Diffusive-\\ dominated\end{tabular}} & \multirow{12}{*}{No} \\ \cline{2-2}
		& 80 \% n\textendash pentane  &                                                                                  &                      \\ \cline{2-2}
		& 90 \% n\textendash pentane  &                                                                                  &                      \\ \cline{2-2}
		& 100 \% n\textendash pentane &                                                                                  &                      \\ \cline{2-2}
		& 70 \% n\textendash heptane  &                                                                                  &                      \\ \cline{2-2}
		& 80 \% n\textendash heptane  &                                                                                  &                      \\ \cline{2-2}
		& 90 \% n\textendash heptane  &                                                                                  &                      \\ \cline{2-2}
		& 100 \% n\textendash heptane &                                                                                  &                      \\ \cline{2-2}
		& 70 \% n\textendash decane   &                                                                                  &                      \\ \cline{2-2}
		& 80 \% n\textendash decane   &                                                                                  &                      \\ \cline{2-2}
		& 90 \% n\textendash decane   &                                                                                  &                      \\ \cline{2-2}
		& 100 \% n\textendash decane  &                                                                                  &                      \\ \cline{2-2} \cline{4-4} 
		& 70 \% n\textendash pentane  &                                                                                  & \multirow{8}{*}{Yes} \\ \cline{2-2}
		& 80 \% n\textendash pentane  &                                                                                  &                      \\ \cline{2-2}
		& 90 \% n\textendash pentane  &                                                                                  &                      \\ \cline{2-2}
		& 100 \% n\textendash pentane &                                                                                  &                      \\ \cline{2-2}
		& 70 \% n\textendash heptane  &                                                                                  &                      \\ \cline{2-2}
		& 80 \% n\textendash heptane  &                                                                                  &                      \\ \cline{2-2}
		& 90 \% n\textendash heptane  &                                                                                  &                      \\ \cline{2-2}
		& 100 \% n\textendash heptane &                                                                                  &                      \\ \cline{2-4} 
		& 100 \% n\textendash heptane & \begin{tabular}[c]{@{}c@{}}Solvent \\ exchange\end{tabular}                      & No                   \\ \hline
		\begin{tabular}[c]{@{}c@{}}100 g/L bitumen \\ in toluene\end{tabular}                     & 100 \% n\textendash heptane & \begin{tabular}[c]{@{}c@{}}Diffusive-\\ dominated\end{tabular}                   & No                   \\ \hline
	\end{tabular}
\end{table}

The yield and size distribution of the asphaltene particles depends on the Hildebrand solubility parameter of the paraffinic solvent and the diffusion coefficient. Hildebrand solubility parameter represents the interaction between compounds. The closer the Hildebrand solubility parameters are, the more likely that the two compounds are miscible \cite{barton2017crc}. The diffusion coefficient affects the duration of the mixing, ultimately affecting the dynamics of asphaltene precipitation \cite{meng2022size}. Therefore, the relative frequency of the PSMPs in different types of precipitants is different. 


\section{Droplet formation and asphaltene precipitation by solvent exchange}

\subsection{Flow rate during the solvent exchange}

\subsubsection{Effects of flow rate on droplet formation}

\citet{zeng2019solvent} proposed a device to control convective mixing: Solvent exchange \cite{zhang2015formation,dyett2017formation,li2018formation,lohse2015surface,wang2021ultrasensitive,zhang2007formation}. Solvent exchange is an effective approach to control the volume, composition, and size distribution of the final surface nanodroplets \cite{you2021tuning}.

\begin{figure}[!ht]
	\centering
	\includegraphics[width=\textwidth]{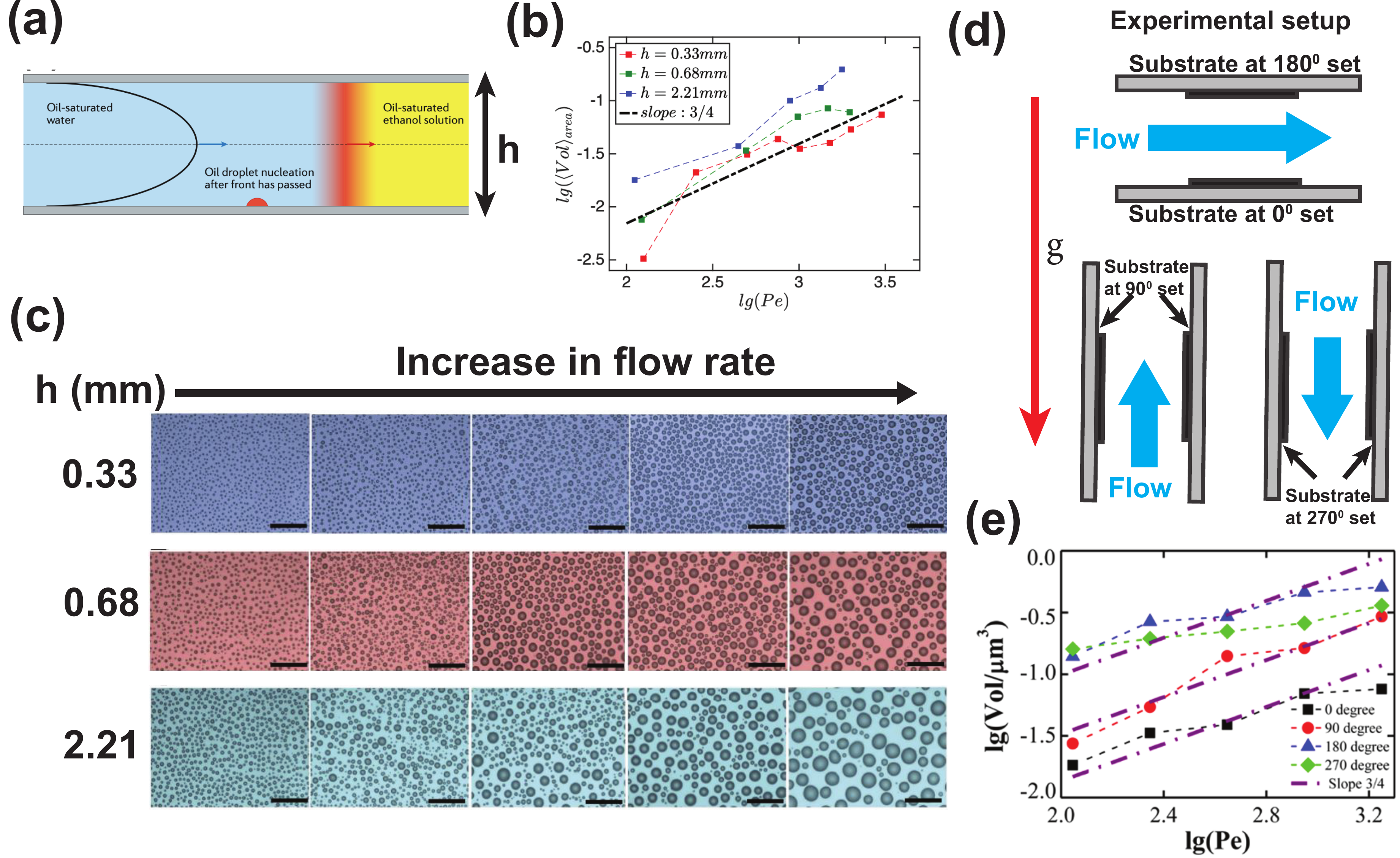}
	\renewcommand{\captionfont}{\linespread{1.6}\normalsize}
	\caption{(a) Sketch for the solvent exchange. (b) Scaling law of P\'{e}clet number and total volume of the surface nanodroplets. (c) Snapshots of the nanodroplets formation under different $Pe$ and channel heights. (d) Different orientations for solvent exchange. (e) Effect of orientation on the total volume of the surface nanodroplets. Panel (a) reproduced with permission from ref \cite{lohse2020physicochemical}. copyright 2020, Springer Nature Limited; Panels (b) and (c) are reproduced with permission from ref \cite{zhang2015formation}. Copyright 2015 National Academy of Sciences; Panels (d) and (e) are reproduced with permission from ref \cite{yu2015gravitational}. Copyright 2015 American Chemical Society.}
	\label{solvent-exchange}
\end{figure}

Figure \ref{TIRF}(c) shows the sketch of solvent exchange.  An increase in flow rate results in an increase in the size of the nanodroplets, as shown in Figure \ref{solvent-exchange}(a) \cite{zhang2015formation}. The experimental data shows a scaling relationship between the total volume of the nanodroplets ($Vol_f$) and the P\'{e}clet number ($Pe$) of the flow(Figure \ref{solvent-exchange}(b)(c)). $Pe$ is defined as the ratio between advective transport rate to diffusive transport rate and can be calculated by:

\begin{equation}
	\label{Pe}
	Pe = \frac{Q}{wD}
\end{equation}
where $Q$ is the flow rate, $w$ is the width of the channel, and $D$ is the diffusion coefficient. The scaling law is:

\begin{equation}
	\label{VOLf}
	Vol_f \sim R_f^3 \sim h^3 Pe^{3/4}
\end{equation}
where $R_f$ is the final radius of the droplets, $h$ is the channel height. The scaling law of $Pe^{3/4}$ is derived according to the diffusive growth of the droplets from the oversaturated droplet liquid transported by an external laminar flow. 
\subsubsection{Effect of flow rate on asphaltene preciptiation}

\begin{figure}
	\centering
	\includegraphics[width=0.8\textwidth]{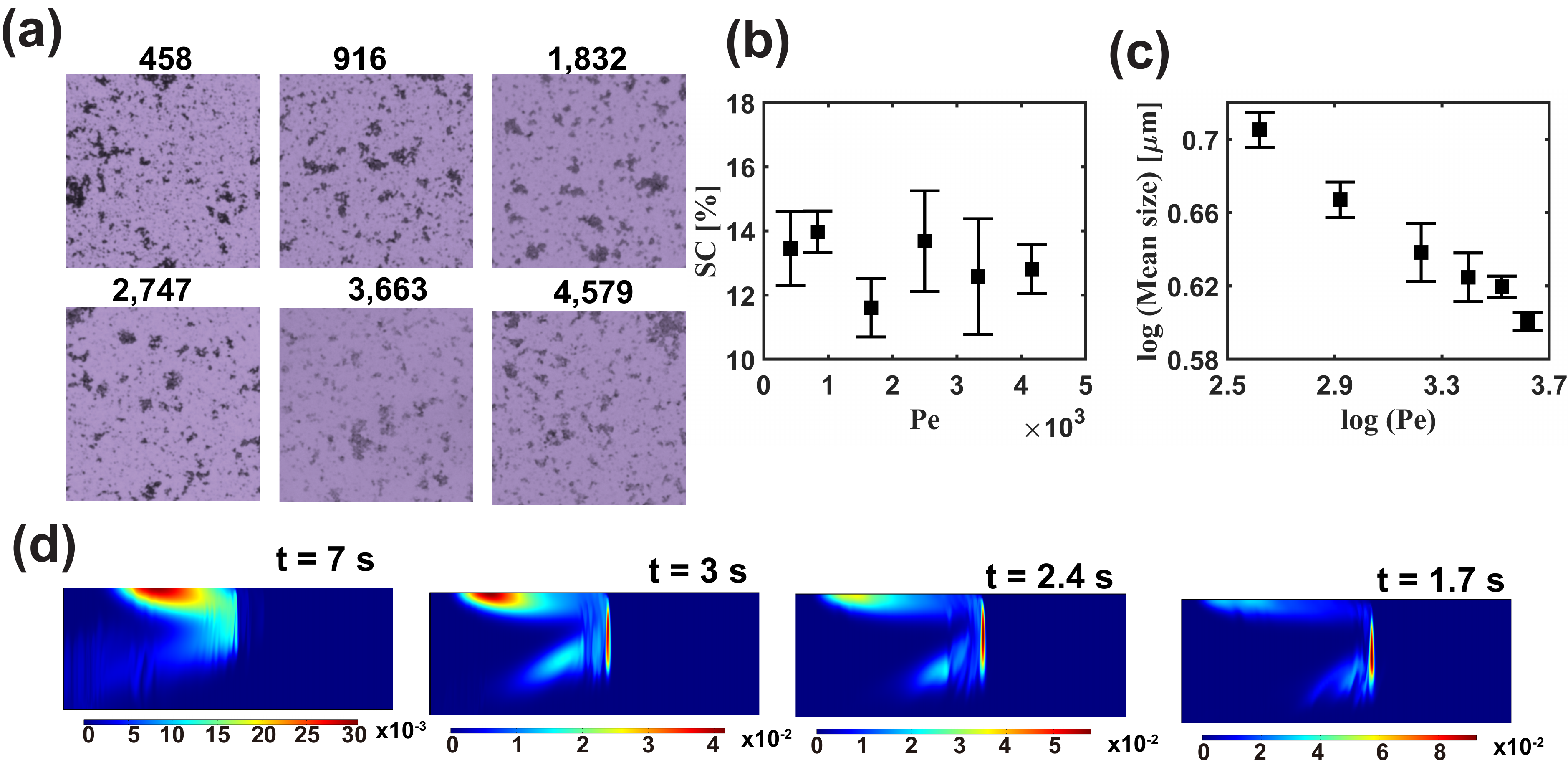}
	\renewcommand{\captionfont}{\linespread{1.6}\normalsize}
	\caption{(a) Snapshots of the asphaltene particles precipitated under different $Pe$. Relationship of (b) surface coverage and (c) size with $Pe$. (d) COMSOL simulation to show the flow profile at different $Pe$ \cite{meng4145264asphaltene}. Panels (a)-(d) are reproduced with permission from ref \cite{meng4145264asphaltene}. Copyright 2020 Elsevier. }
	\label{flow rate}
\end{figure}

Inspired by the influence of the mixing conditions in the liquid\textendash liquid phase separation process, \citet{meng4145264asphaltene} used microfluidic devices to control the mixing of a model oil and paraffinic solvents to study dilution\textendash induced asphaltene precipitation. Asphaltene solution prefills the channel initially. Then precipitant is injected to displace the asphaltene solution, and the asphaltene precipitation happens simultaneously in this displacing process \cite{zhang2015formation}. Different from the liquid\textendash liquid phase separation, the scaling law of total volume to $Pe$ in Equation (\ref{VOLf}) does not hold for the system of asphaltene precipitation, as shown in Figure \ref{flow rate}(a)(b). COMSOL shows that n\textendash heptane is concentrated at the top of the channel due to the large density difference between n\textendash heptane and toluene. n\textendash Heptane concentration at the bottom is always very low near the substrate due to the density difference between n-heptane and toluene, and this may be the reason that surface coverage of the asphaltene particles is not influenced by $Pe$ (Figure \ref{flow rate}(d)). 

In addition, the size of the asphaltene particles decreases with the increase of $Pe$ (Figure \ref{flow rate}(c)). \citet{abbas2021deposition} also found that the size of the asphaltene particles deposited on the surface decreases with the increase of shear rates due to the breakage of asphaltene aggregates.

\citet{elkhatib2019nanoscale} split the asphaltene particles into four classes, including nano-aggregates (1.5 \textendash 4 nm), small clusters (SCs, 4 \textendash 10 nm), medium clusters (MCs, 10 \textendash 20 nm), large clusters (LCs, 20 \textendash 100 nm), and extra\textendash large clusters (XLCs, $>$ 100 nm). Among them, LCs and XLCs are more easily to be removed from the surface by inertia forces originating from the flow. \citet{sarsito2021suppression} found the deposition of asphaltene on the different surfaces may be different due to the hydrophobicity. Therefore, changes in flow rate affect not only asphaltene precipitation but also asphaltene deposition.

\subsection{Effects of dimension and orientation of the channel}

\subsubsection{Effects on droplet formation}

From Equation (\ref{VOLf}), for the same $Pe$, the total volume of the nanodroplets increases with the channel height. When the channel height is larger, the distance from the substrate to the position of the maximum flow velocity is longer due to the parabolic flow. Therefore, the duration of the oversaturation pulse is longer, resulting in a larger volume of nanodroplets. \citet{meng4145264asphaltene} observed a similar phenomenon in the asphaltene system. The surface coverage increases with the Rayleigh number ($Ra$), which is calculated by Ra $\sim h^3$ (Figure \ref{orientation}(a)). The COMSOL simulation visualized enhanced mixing from a larger channel height (Figure \ref{orientation}(b)(c)).

When the parabolic shape of the flow deforms, the distance between the walls to the position of the maximum flow rate varies when the density difference of solutions of A and B is significant enough. This distance variance only happens when the flow direction is perpendicular to gravity. The distance between the wall and the maximum flow point is not affected by gravity when the flow direction is parallel to the gravity direction, as shown in Figure \ref{solvent-exchange}(d) \cite{yu2015gravitational}.

\citet{yu2015gravitational} placed the device of solvent exchange horizontally and vertically, and they found that the scaling law of $Pe$ and $Vol_f$ is still valid. However, under the conditions of the same $Pe$ and channel height, due to the deformation of the parabolic profile, the volume of the nanodroplets of the 180$^\circ$ was higher than 180$^\circ$ and 0$^\circ$ when the density of solution B was higher than solution A (Figure \ref{solvent-exchange}(e)). In addition to the volume, the size distribution of the nanodroplets was also changed by varying the orientation \cite{yu2015gravitational}. 

For the horizontally placed device, if the deformation of the parabolic profile happens can be determined by Archimedes' number ($Ar$) of the system, which is calculated by:

\begin{equation}
	\label{Ar}
	Ar = \frac{gh^3}{\nu^2} \frac{\Delta \rho}{\rho}
\end{equation}
where $g$ is the gravity constant, $\nu$ is the kinematic viscosity, $\rho$, and $\Delta \rho$ are the density of solution B and the density difference between solutions of A and B, respectively. Gravitational effect plays a predominated role when $Ar >> 1$. Therefore, the orientation effect only occurs when the channel height is high enough.

\citet{zhang2015formation} found that the increase in height led to an increase in the duration of the oversaturation pulse and the emergence of convective rolls. The convective rolls enhanced the mixing between solutions of A and B, leading to the increase in the volume of the nanodroplets. However, the scaling law between $Pe$ and total volume is still valid because the convective rolls only affect the prefactors in Equation (\ref{VOLf}).

\subsubsection{Effects on asphaltene precipitation} 
\begin{figure}[!ht]
	\centering
	\includegraphics[width=\textwidth]{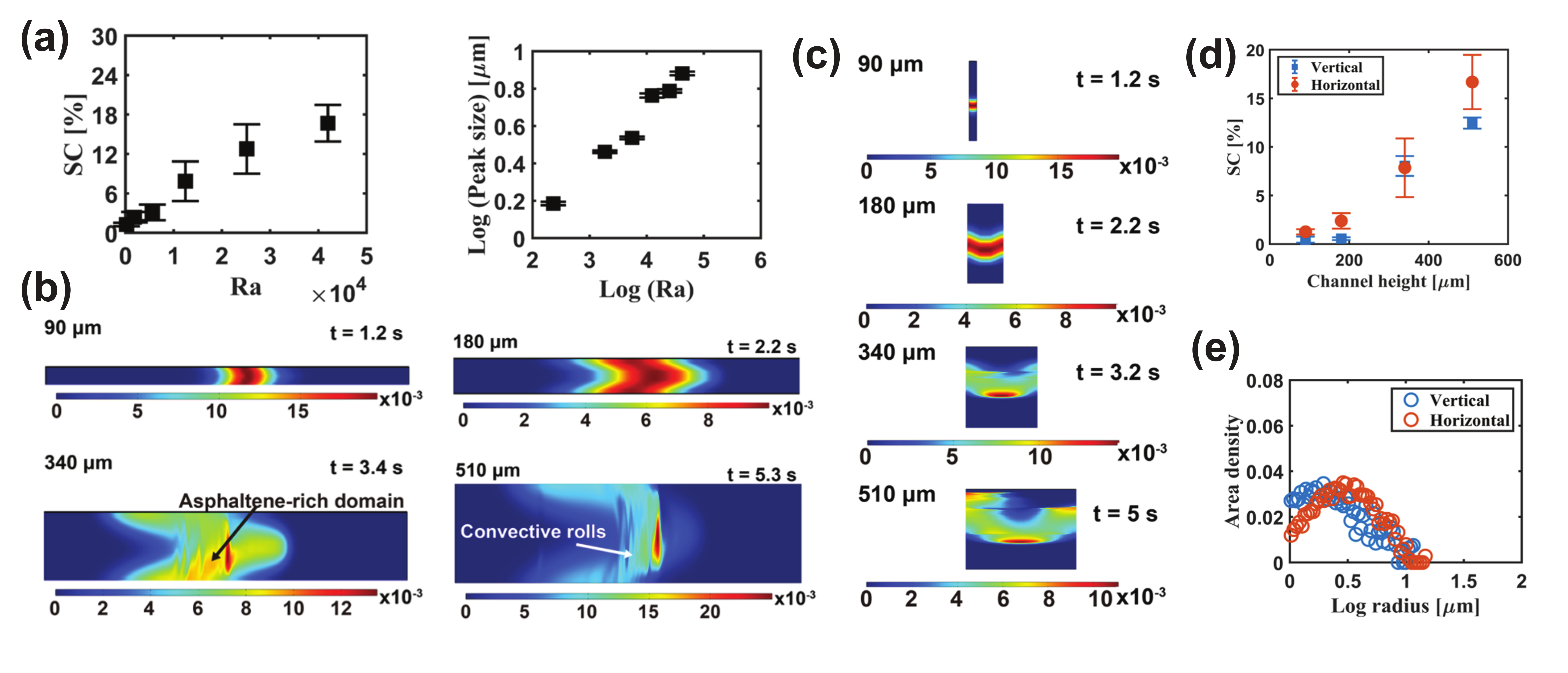}
	\renewcommand{\captionfont}{\linespread{1.6}\normalsize}
	\caption{(a) Effect of channel heights on surface coverage and size of asphaltene particles. COMSOL simulation to show the flow profile in different channel heights of (b) horizontally and (c) vertically placed devices. Effect of orientation on (d) surface coverage and (e) size distribution of asphaltene particles \cite{meng4145264asphaltene}. Panels (a)-(e) are reproduced with permission from ref \cite{meng4145264asphaltene}. Copyright 2020 Elsevier.}
	\label{orientation}
\end{figure}

COMSOL simulation visualized that the convective rolls also exist in the asphaltene system. The mixing is thus enhanced, and an asphaltene\textendash rich domain appears at the tail of the mixing front, as shown in Figure \ref{orientation}(b). The mass transfer of asphaltene from the flow to the substrate is enhanced due to the longer transfer time and concentration gradient. Therefore, compared with the vertically placed device, the horizontal group shows a higher surface coverage and larger size, as shown in Figure \ref{orientation}(d)(e) \cite{meng4145264asphaltene}.

\section{Challenges and Future Perspectives}
In analogy to the formation of nanodroplets induced by the Ouzo effect, a recent study has revealed the effect of mixing dynamics on the asphaltene precipitation. The previous studies on nanodroplet formation can be used as references for the study of asphaltene precipitation. However, the two processes are not the same because asphaltene is more complex and heterogeneity in molecular structures. 

Researchers have demonstrated that mixing conditions affect asphaltene precipitation dynamics while solvent consumption remains constant. Nevertheless, a systematic correlation does not exist. Despite its importance, this aspect of the research has received insufficient attention. Understanding the fundamental insights helps to advance microscopy, microfluidics, and computational fluid dynamics. Asphaltene precipitation can be visualized at micro and nanometers. A microfluidic device can be designed to control mixing conditions, and simulation software can be used to characterize flow conditions. The following directions can be considered by researchers in the future. 

\begin{enumerate}
	\item Understand the effects of molecular structures of the asphaltene on the precipitation under different mixing conditions. Asphaltene of different molecule structures and molecular weight differ in their interactions with each other and the surrounding solvent. It is impossible to isolate and obtain asphaltenes of the same molecular structures. But it will be insightful and feasible to study precipitations of several fractions where the asphaltenes with similar sizes are grouped together. The more quantitative study will help to understand the intermolecular forces between asphaltene molecules to enhance the development of more reliable and predictive models.
	
	\item computational fluid dynamics (CFD) tools can be utilized to correlate the mixing conditions with the yield and size distribution of asphaltene particles. The current study of the effects of mixing dynamics on asphaltene precipitation is still preliminary because of limited mixing regimes. The strong light-adsorbing nature of asphaltenes also limits the in-situ experimental study of the effects from the mixing dynamics. The flow and composition of liquids may be visualized in CFD simulations as the physical properties of the solutions are defined, including density, viscosity, and temperature. Correlating the CFD results with the experimental asphaltene yield and particle size distribution may help further understanding the effects of mixing on asphaltene precipitation.
	
	\item Molecular spectroscopes with even higher spatial resolution can be used to characterize the finer morphology of asphaltene precipitates of known composition. The process of asphaltene precipitation from nano-aggregates to micron-sized particles in the current study is still challenging to be followed in time. The particle size obtained by terminating the aggregation process at the early phase separation stage may be below the visible range of the conventional optical microscope. With the advancement of super-resolution microscopic technology, it may be possible to obtain high-resolution morphology and chemical composition simultaneously. The morphology may be therefore correlated with the molecular structures of asphaltenes and the aggregation process of asphaltene from nano-aggregates to microparticles.
	
	\item Removal of the good solvent from the mixture to induce asphaltene precipitation. Preferential evaporation or dissolution of the good solvent from drops of a ternary mixture can also trigger the droplet formation \cite{tan2016,tan2017,tan2019}, following the same mechanism for standard ouzo effect from the addition of the poor solvent, but in a reverse direction by reducing the ratio of the good solvent. Extremely rich transport phenomena driven Marangoni effect may occur during the evaporation or dissolution of the ouzo drops. Considering the drastically different interfacial tension and diffusion constants of the composition, it will be
	intriguing to study the effects of the selective solvent removal on asphaltene precipitation and explore how far the understanding based on well-defined aqueous systems can be applied to multicomponent organic mixtures of high complexity. Such a study may not be closely related to large-scale applications, but it contributes to a fundamental understand of phase separation in complex systems.

\end{enumerate}

\cleardoublepage
\section{Conclusions}

Phase separation induced by the ouzo effect is a good analogy for understanding asphaltene precipitation. The droplet size from phase separation induced by the ouzo effect is determined by the chemical composition and mixing conditions. By leveraging advanced microfluidic techniques, the influence of mixing conditions on the ouzo effect can be well controlled.
In parallel, asphaltene precipitation is a complex process influenced by thermodynamic factors, including the S/B ratio and type of precipitant. The use of microfluidic devices has brought about a turning point for understanding the effects of mixing dynamics \cite{sharma2022perspectives}. Primary submicron particles form ubiquitously from mixing with the diluent under a wide range of mixing conditions. Without changing the use of solvent, the total yield of asphaltene and particle size can be adjusted by changing the mixing conditions. New techniques such as confocal and TIRF are complemented with computational fluid dynamics simulations to quantitatively understand the effects of mixing dynamics on asphaltene precipitation. 

Many physico and hydrodynamical features are shared by the ouzo effect or nanoprecipitation in multicomponent aqueous systems and diluent\textendash induced asphaltene precipitation. We hope that the comparison in this review will inspire more in\textendash depth understanding of both asphaltene precipitation and the ouzo effect in aqueous systems.
\cleardoublepage
\section*{Declaration of Competing Interests}
The authors declare that they have no known competing financial interests or personal relationships that could have appeared to influence the work reported in this paper.

\section*{Biographies}
\noindent \textbf{Jia Meng} received his Ph.D. degree in Chemical and Materials Engineering under the supervision of Prof. Xuehua Zhang from the University of Alberta, Edmonton, AB, Canada. His areas of research include soft matter and interface, microfluidic technology, and oil sands extraction.

\noindent \textbf{Somasekhara Goud Sontti} is a  Postdoctoral researcher working in the Department of Chemical and Materials Engineering under the supervision of Prof. Xuehua Zhang at the University of Alberta, Canada. Dr. Sontti obtained his Ph.D. and M. Tech degrees in Chemical Engineering from the Indian Institute of Technology Kharagpur and the Indian Institute of Technology Guwahati, India. His research interest includes the broad area of multiphase flow, microfluidics, computational fluid dynamics, and slurry transport. 

\noindent \textbf{Xuehua Zhang} is a Professor and Canada Research Chair (Tier 1) at the Department of Chemical and Materials Engineering, University of Alberta, Canada.  Before, she worked at Australian National University, University of Melbourne, and RMIT University, Australia. The present research interests of Dr. Zhang range from microscopic bubbles and drops in separation technology and water treatment to evaporation and chemical reactions of droplets for ultrasensitive detection.  She conducts both fundamental and applied research and combines experimental, theoretical, and numerical methods. 
\section*{Acknowledgement}
The authors are genuinely grateful for the inspiring discussions with Xiaoli Tan and Murray R Gary over the topic over the years. The authors acknowledge the support from Natural Science and Engineering Research Council of Canada (NSERC)\textendash Collaborative Research and Development Grants, the Canada Research Chair Program, and from Canada Foundation for Innovation, John R. Evans Leaders Fund. This work is partially supported by the Canada Research Chairs program.

\cleardoublepage
\bibliography{literature}

\end{document}